# 3
# Design principles for super selectivity using multivalent interactions

*Tine Curk, Jure Dobnikar, Daan Frenkel*

## 3.1
## Introduction

Multivalent particles have the ability to form multiple bonds to a substrate. Hence, a multivalent interaction can be strong, even if the individual bonds are weak. However, much more interestingly, multivalency greatly increases the sensitivity of the particle-substrate interaction to external conditions, resulting in an ultra-sensitive and highly non-linear dependence of the binding strength on parameters such as temperature, pH or receptor concentration.

In this chapter we focus on super selectivity: the high sensitivity of the strength of multivalent binding to the number of accessible binding sites on the target surface (see the schematic drawing in Figure 3.1). For example, the docking of a multivalent particle on a cell-surface can be very sensitive (super selective) to the concentration of the receptors to which the multiple ligands can bind.

We present a theoretical analysis of systems of multivalent particles and describe the mechanism by which multivalency leads to super selectivity. We introduce a simple analytical model that allows us to predict the overall strength of interactions based on physiochemical characteristics of multivalent binders. Finally, we formulate a set of simple design rules for multivalent interactions that yield optimal selectivity.

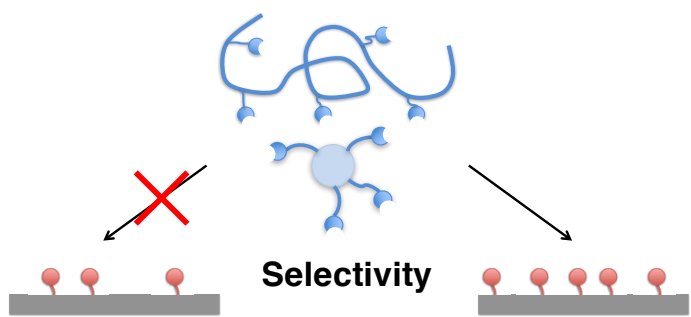

**Figure 3.1** Selectivity denotes the ability of multivalent entities to distinguish between substrates depending on the surface density of binding sites.

### 3.1.1
**Background: Ultra-sensitive response**

Many processes in biology depend ultra-sensitively on variations in one or more of the parameters that control the process. Such ultra-sensitivity manifests itself as an almost switch-like, sigmoidal change in the 'output' when the control parameter crosses a threshold value. Understanding such switch-like behaviour is obviously important to understand many regulatory processes in living systems, but such understanding will also help us design synthetic systems that combine weak supramolecular interactions with high selectivity.

The best known example of ultra-sensitivity dates back to Hill who, in the beginning of the twentieth century, studied the binding of oxygen to haemoglobin. He found the that the relation between bound oxygen and partial pressure was sigmoidal [1]. Today this phenomenon is explained in terms of allosteric cooperativity whereby the 4 binding sites on haemoglobin do not act independently but are 'cooperative', i.e. binding of the first oxygen molecule increases the probability that the second oxygen molecule will bind. Hence, haemoglobin is likely to be either fully loaded with oxygen or empty, which makes haemoglobin an efficient transporter of oxygen between lungs and peripheral tissues. Other examples of ultra-sensitivity include the switch-like response of bacterial motors [2], or the switch-like behaviour in gene regulation due to positive feedback loops in nucleosome modification [3]. For more information on this broad topic, the reader is referred to a review by Ferrell [4, 5, 6] and references therein.

Ultra-sensitive response is usually characterised by a so-called Hill curve:

$$\text{Output} = \frac{\text{Input}^n}{K^n + \text{Input}^n} , \tag{3.1}$$

where the Hill coefficient $n$ quantifies the degree of cooperativity of the process: the higher the Hill coefficient, the more sensitive the response [1)].

Due to cooperativity, blocks that, individually, have limited selectivity can form units that interact selectively. For example, DNA base pairing is highly specific, even though underlying interactions (hydrogen bonding and base-stacking) are not. Multivalent (or polyvalent) interactions can also lead to an ultra-sensitive response, for example, the aggregation of multivalent DNA-coated colloids depends sensitively on temperature [7]. Moreover, ligand-receptor or antibody-antigen interactions, are very sensitive to temperature, but also to ion concentration and pH. Internal protein interactions are also multivalent: protein folding and unfolding depend critically on temperature and other external conditions. The functioning of the biochemical machinery in cells relies (mostly) on multivalent supra-molecular interactions. These interactions are very sensitive to external conditions which helps explain why the properties of living matter (cells, tissues) are very sensitive to temperature, while those of 'formerly living' matter (say, a piece of wood) are not.

---

1) In order to make this chapter accessible to a broad audience, we keep mathematical expressions in the main text to an absolute minimum: well known relations, such as the Langmuir isotherm, and our final design principles are included because they are needed to understand super selectivity. However, all other equations and mathematical derivations are enclosed in boxes for the aficionado: readers less interested in the mathematical background can skip these without risk.

---

**Multivalent interactions: Why so sensitive ?**

Imagine two multivalent entities at a fixed distance that are connected by a number of bonds (say $k$). The two entities can dissociate only when all $k$ bonds are broken. We denote the probability that an individual bond is broken by $p_1^{\text{unbound}}$ and the probability that all $k$ bonds are broken by $p_k^{\text{unbound}}$. If different bonds do not influence each other, the probability of unbinding is

$$p_k^{\text{unbound}} \sim \left(p_1^{\text{unbound}}\right)^k . \tag{3.2}$$

Note that for large 'valencies' $k$, the relation between $p_1^{\text{unbound}}$ and $p_k^{\text{unbound}}$ is highly non-linear. In fact, the expression for the ratio between probabilities $p_k^{\text{unbound}}/p_k^{\text{bound}}$ can be written in a form reminiscent of the Hill equation:

$$\frac{p_k^{\text{unbound}}}{p_k^{\text{bound}}} \sim \frac{\left(p_1^{\text{unbound}}\right)^k}{1 - \left(p_1^{\text{unbound}}\right)^k} , \tag{3.3}$$

where the exponent $k$ plays a role similar to that of the Hill coefficient (Eq. (3.1)). The probability of a single bond spontaneously breaking $p_1^{\text{unbound}}$ will depend not only on control parameters such as bond strength, temperature, pH of the solution etc., but also on the number of possible bonding arrangements. Clearly, the unbinding probability, Eq. (3.2), tends to be very sensitive to any parameter that influences $p_1^{\text{unbound}}$. This example illustrates the physical origins of ultra-sensitive response in multivalent interactions. We shall see below that competition between different bonds modifies the response but retains ultra-sensitivity.

---

In what follows, we focus on the ultra-sensitivity of multivalent interactions to the density of 'receptors' on the substrate surface. In particular, we will derive expressions that show how the binding strength of a multivalent entity (say a ligand-decorated nanoparticle or a multivalent polymer) to a substrate changes with the concentration of receptors [2] on the substrate surface (see Figure 3.2). It will turn out that multivalent interactions can be designed such that they result in an almost step-like switch from unbound to bound as the receptor concentration exceeds a well-defined threshold value. In the remainder of this chapter, we will use the term 'super selectivity' to denote this kind of sharp response.

---

[2] A brief comment on the use of terminology: we make liberal use of the terms 'ligand' and 'receptor' with which we shall denote individual binding partners. 'Receptors' will be found on the substrate surface whilst individual 'ligands' are attached to the multivalent entity (say, a nano-particle) that binds to the substrate, shown in Figures 3.2 and 3.3. We use the term 'multivalent entity' to denote any moiety that is able to form multiple bonds. The term 'binding site' always denotes an individual monovalent interaction site, equivalent to a single 'ligand' or 'receptor'.

**4**

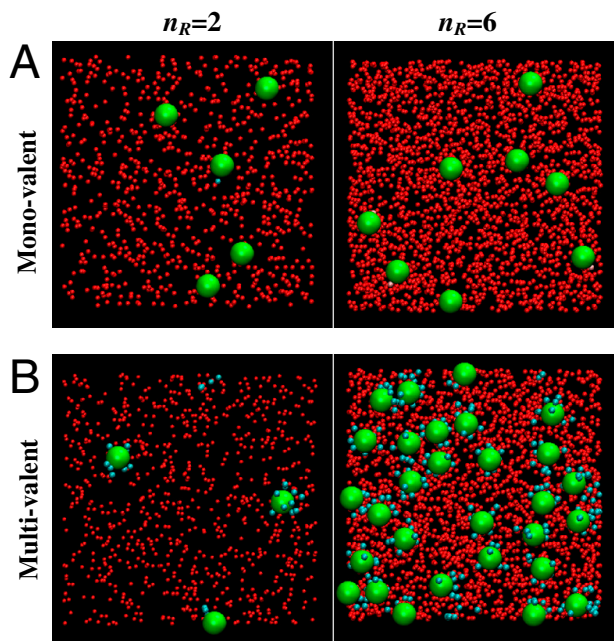

**Figure 3.2** Simulation snapshots comparing the targeting selectivity of monovalent and multivalent guest nano-particles. We compare the adsorption onto two host surfaces with receptor concentrations ($n_R$) that differ by a factor of three. (A) The monovalent guests provide little selectivity: increasing by three times the receptor coverage just increases the average number of bound guests by 1.8 (i.e., from 5.4 to 9.7 bound particles in average). (B) The multivalent nano-particles behave super selectively: an increase of the receptor coverage by a factor three causes a 10-fold increase in the average number of adsorbed particles. The multivalent guests have ten ligands per particle. The individual bonds of the multivalent case (B) are weaker than those in the monovalent case (A). Figure reproduced with permission from Ref. [8].

The remainder of this chapter is structured as follows: First, we show how the description of simple chemical equilibria and Langmuir adsorption can be extended to multivalent interactions. We then discuss the conditions under which super selectivity appears and formulate simple design principles to achieve super selectivity. We include an appendix where we discuss how, in simple cases, our approach reduces to the widely used 'effective molarity' picture.

## 3.2
## Super selectivity: an emergent property of multivalency

We first focus on a prototypical system of multivalent particles in solution that can adsorb to a receptor-decorated surface (see Figure 3.2). For simplicity, we assume that the surface is flat and much larger than the multivalent particles. Furthermore, we assume that these particles are larger than the surface receptors such that each particle can attach to many receptors sites simultaneously. Adsorption of particles is governed by the well-known Langmuir isotherm which states that the fraction of the surface occupied by particles is

$$\theta = \frac{\rho K_A^{av}}{1 + \rho K_A^{av}}, \qquad (3.4)$$

with $\rho$ the molar concentration of particles in solution [3], $K_A^{av}$ is the equilibrium avidity association constant of particles adsorbing to a surface. Note that $K_A^{av}$ is different from the affinity equilibrium constant $K_A$ which specifies chemical equilibria of individual ligand-receptor binding. Avidity (functional affinity) is the accumulated strength of multiple affinities [9].

We aim to understand how the overall avidity constant $K_A^{av}$ depends on the properties of the system, i.e. individual bond affinities $K_A$ [4], the ligand valency $k$ and number of receptors $n_R$. The avidity constant includes all possible bound states, and is written as a sum over bonds

$$K_A^{av} = \Omega_1 K_A + \Omega_2 K_A K_{intra} + \Omega_3 K_A K_{intra}^2 + ... \qquad (3.5)$$

The first term on the right hand side takes into account all states with a single formed bond, the second term represents all doubly bound states, the third term triply bound states, etc. $K_{intra}$ is a constant specifying the internal equilibrium between singly and doubly bonded states. We have assumed that individual bonds form independently and $K_{intra}$ is a constant, i.e. we ignore (allosteric) cooperative effects. We do this to clearly distinguish multivalent effects (the subject of this chapter) from cooperative effects [10] [5]

$\Omega_i$ is the degeneracy pre-factor, it measures the number of ways in which $i$ bonds can be formed between two multivalent entities, see Figure 3.3 for representative cartoons. Degeneracy $\Omega$ is often labelled as a 'statistical pre-factor' which denotes something that should be included for rigour but is otherwise not essential. However, as we will show, it is precisely this degeneracy that gives rise to super selectivity. The focus of the majority of theoretical papers [9, 11, 12, 13, 14] is on the calculation of the internal equilibrium constant $K_{intra}$. Here, instead, we focus on the degeneracy $\Omega$. We will simply assume that $K_{intra}$ is (or can be) known.

The degeneracy $\Omega$ depends on the spatial arrangement of both ligands and receptors. However, it is instructive to consider first the binding of flexible ligands, where all $k$ ligands on a particle can bind to $n_R$ receptors (Figure 3.3B). In this case the degeneracy given by Eq. (3.6) becomes a very steep and non-linear function of $k$ and $n_R$. This form was first considered by Kitov and Bundle [15] and has been applied, among others, to super-selective targeting [8] and modelling the adhesion of influenza virus [16].

---

3) For non-ideal solutions the density $\rho$ in the Langmuir isotherm (Eq. (3.4)) should be replaced by the fugacity.
4) $K_A$ is the association equilibrium constant between a monovalent particle (a single ligand attached to a particle) and a single receptor, we assume it can be determined experimentally
5) Some authors [11] use the term 'chelate cooperativity' to denote multivalent effects.

> **Degeneracy $\Omega$**
>
> In the 'flexible' binding case where each of the $k$ ligands can bind to every one of the $n_R$ receptors that is is available, the number of ways (degeneracy) to form $i$ bonds is
>
> $$\Omega_i = \binom{n_R}{i}\binom{k}{i} i! = \frac{n_R!\, k!}{(n_R - i)!\, (k - i)!\, i!}\;. \tag{3.6}$$
>
> We need to choose $i$ ligands out of $k$ and choose $i$ receptors out of $n_R$, then there are $i!$ (that is $i$ factorial) ways of binding the chosen ligands/receptors together.

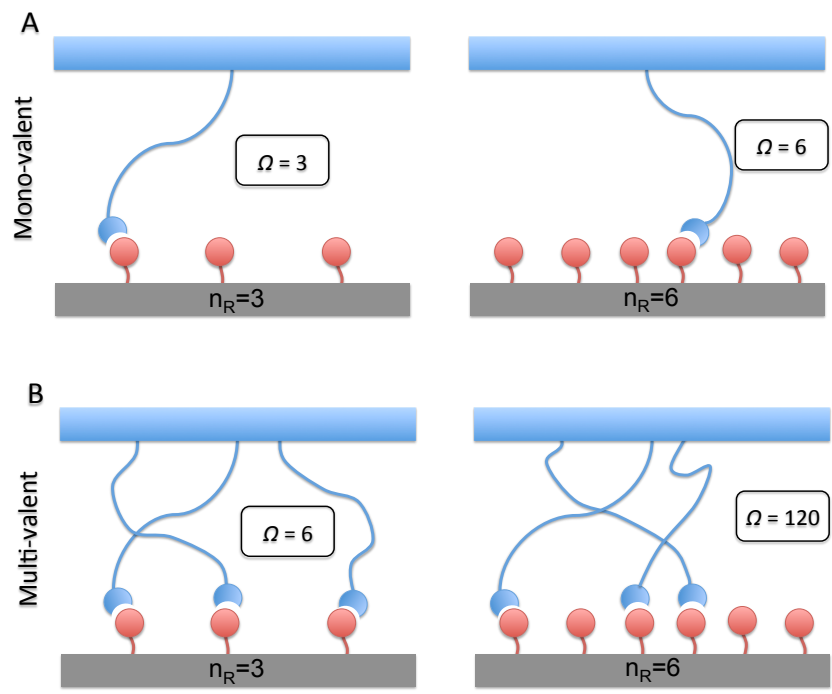

**Figure 3.3** Entropic origin of super selectivity. The cartoons give a schematic representation of the simulation snapshots in Figure 3.2. The pictures show the binding of mono-valent (A) and multi-valent (B) entities (represented as a bar with attached flexible ligands). Receptors are shown as spheres tethered to the bottom surface. The left panels show a low receptor density ($n_R = 3$) and the panels on the right show a receptor density that is twice as high. In the mono-valent case the number of distinct ways ($\Omega$) to link ligands and receptors grows linearly with the number of receptors $n_R$, while the multi-valent case show a highly non-linear response: changing $n_R$ from 3 to 6 increases $\Omega$ by a factor of 20. In general, the number of binding combinations (degeneracy) $\Omega$ is calculated using Eq. (3.6).

A low fraction of bound receptors in the system can arise either because the number of receptors is greater than the number of available ligands: $n_R \gg k$ or when individual bonds are weak: $K_{intra} \ll 1/k$ [6]. In this case the avidity constant (Eq. (3.5)), using degeneracy (Eq. (3.6)), can be rewritten [7] to yield a simple form:

$$K_A^{av} \approx \frac{K_A}{K_{intra}} \left[ (1 + n_R K_{intra})^k - 1 \right], \qquad (3.7)$$

where, as before, $K_A$ is the monomeric single-bond affinity constant, $K_{intra}$ the internal association constant, and $n_R$ and $k$ are the number of receptors and ligands respectively. For our purpose it is important to note that for multivalent binding ($k > 1$), $K_A^{av}$ is a steep, non-linear function of $n_R$ (see Figure 3.4).

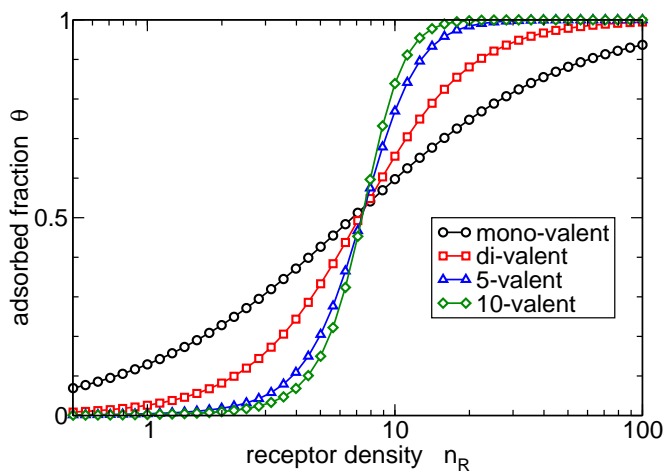

**Figure 3.4** Adsorption profile of multi-valent particles computed using Eqs. (3.4) and (3.7). Monovalent adsorption (black circles) $k = 1$ yields the familiar Langmuir isotherm. In contrast, multivalent particles display a steep, sigmoidal response. In the case shown in the Figure, we have chosen the dimensionless activity in solution to be $z \equiv \rho \frac{K_A}{K_{intra}} = 0.001$, the binding affinity of individual bonds decreases as the valency increases from mono-valent to 10-valent: $\log(K_{intra}) = 5, 1.5, -1, -2$, such that the overall avidity $K_A^{av}$ at $50\%$ bound fraction ($\theta = 0.5$) is kept constant for all valencies.

Eq. (3.7) could have also been obtained directly by reasoning that for non-saturated receptors (fraction of bound receptors is low), competition for the same receptor can be ignored. Each of the $k$ ligands can then bind independently to any of the $n_R$ receptors with an equilibrium constant $K_{intra}$ (weight $n_R K_{intra}$). Alternatively, the ligand is unbound (weight 1). Hence, for systems with a low fraction of bound receptors, the factor $(1+n_R K_{intra})^k$ accounts (approximately) for all possible states. Furthermore, we subtract 1 because we use the convention that at least a single bond needs to be formed for the multivalent particle to be considered bound. The avidity constant has units of inverse molar concentration. To obtain the correct limiting behaviour in the limit $k = 1$, where $K_A^{av} = n_R K_A$, we must multiply expression in square brackets by $\frac{K_A}{K_{intra}}$.

---

6) The largest term in (3.5) is obtained by $\Omega(i) K_{intra}^i \approx \Omega(i+1) K_{intra}^{i+1}$, which results in $K_{intra} \approx \frac{i}{(k-i)(n_R-i)}$. If the bonds are sufficiently weak: $K_{intra} < 1/k$, the largest term will always arise when the fraction of occupied receptors is low $\frac{i}{n_R} < 0.5$.

7) Eqs. (3.5) becomes a binomial expansion series that we can sum [8].

We note that the ratio $\frac{K_A}{K_{intra}} = v_{eff}$ has the dimension of an effective volume $v_{eff}$. The form of Eq. (3.7) suggest that we can view the multivalent particle adsorption as a two-step process. First, the particle adsorbs from the solution to the surface and comes into a position to start forming bonds, the equilibrium constant of this process is given by the ratio $\frac{K_A}{K_{intra}}$. Once the particle is in this position, all of the $k$ ligands can independently form bonds with surface receptors.

In the monovalent case ($k = 1$) the avidity constant reduces to $K_A^{av} = n_R K_A$ and the standard Langmuir isotherm is obtained. Furthermore, expanding Eq. (3.7) in a binomial series and using a maximum term approximation we can insert the maximum term in Eq. (3.4) and obtain the phenomenological Hill equation (Eq. (3.1)). In the case of very strong individual bonds ($n_R K_{intra} \gg 1$) virtually all $k$ bonds are formed and the avidity becomes $K_A^{av} \approx n_R^k K_A K_{intra}^{k-1}$ [8].

> **Notation**
>
> In this chapter we choose to work with equilibrium constants and densities as our quantities of choice. However, in earlier work we used a notation based on statistical mechanics. In that case, the central quantities are binding free energies and partition functions. This box provides a translation cheat-sheet between the chemical and statistical mechanical language:
>
> - Gibbs free energy of forming the first bond: $e^{-\beta \Delta G} = K_A \rho_0$
> - Binding free energy of subsequent bonds: $e^{-\beta f} = K_{intra}$
> - Bound state partition function: $q_b = K_A^{av} \frac{K_{intra}}{K_A}$
> - Dimensionless activity of multivalent ligands in solution: $z = \rho \frac{K_A}{K_{intra}}$
>
> where $\rho_0 = 1\text{M}$ is the standard concentration, $\beta = 1/k_B T$ is the inverse of temperature $T$ and $k_B$ the Boltzmann constant. Using these identifications, we can rewrite the surface coverage Eq. (3.4) as
>
> $$\theta = \frac{z q_b}{1 + z q_b} , \qquad (3.8)$$
>
> where the bound partition function is given by
>
> $$q_b = \left(1 + n_R e^{-\beta f}\right)^k - 1 . \qquad (3.9)$$
>
> This dimensionless notation was used in Refs. [8, 17, 18, 19]. We will use it below when formulating general design principles.

---

8) this holds for $n_R \gg k$ when Eq. (3.7) is applicable even for strong bonds, in general (using Eq. (3.5)) the expression would be $K_A^{av} = \frac{n_R!}{(n_R - k)!} K_A K_{intra}^{k-1}$

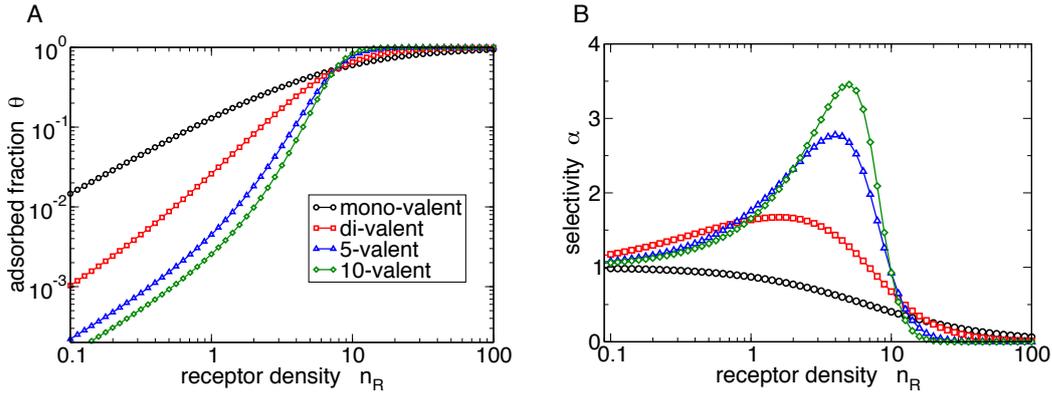

**Figure 3.5** Selectivity. **A)** shows the log-log plot of Figure 3.4, and **B)** shows its slope, i.e. the selectivity $\alpha = \frac{d \log \theta}{d \log n_R}$. We observe that selectivity is typically less than one for mono-valent particles indicating at most linear response. Multi-valent particles, on the other hand, exhibit a region with values of $\alpha$ significantly greater than one, thus demonstrating that the number of adsorbed ligands increases faster than linearly with the receptor concentration: in this regime, the system is super selective.

We have shown how combinatorial entropy (also called 'avidity entropy' [9]) gives rise to sharp switching behaviour upon a change in receptor concentration $n_R$ (Figure 3.4). Next, we introduce a measure of the sensitivity of the binding of multi-valent particles to the surface concentration of receptors:

$$\alpha = \frac{d \log \theta}{d \log n_R} \ . \tag{3.10}$$

$\alpha$ is the slope of the adsorption profile in a log-log plot (see Figure 3.5). For mono-valent binding the selectivity $\alpha$ is never larger than one, while in the multi-valent case the selectivity can reach values greater than one, indicating a supra-linear response. Note that for low surface coverage ($\rho K_A^{av} \ll 1$) the selectivity $\alpha$ is equivalent to the effective Hill-coefficient $n$ from Eq. (3.1). However, because we consider all terms (all possible number of bonds) in calculating avidity (Eq. (3.7)), $\alpha$ is not a constant. At very low receptor concentrations the avidity shows a linear dependence on $n_R$, and $\alpha \approx 1$ [9]. Selectivity then grows with increasing receptor concentration $n_R$ until reaching a peak just before the saturation of the surface ($\rho K_A^{av} \approx 1$). We refer to the region with $\alpha > 1$ as the 'super-selective' region. In this region, a small change in the receptor density $n_R$ causes a faster-than-linear change in adsorption $\theta$.

---

9) At low receptor concentration for $n_R k K_{intra} \ll 1$, expanding Eq. (3.7) to first order we obtain $K_A^{av} \approx n_R k K_A$.

## 3.3
## Multivalent polymer adsorption

To validate the model for super-selective adsorption described above, we now compare its predictions with experimental data on polymer adsorption. Multivalent glycopolymers have been used as selective probes for protein-carbohydrate interactions in a biochemical setting [20, 21, 22]. More recently, super-selective targeting was demonstrated in a synthetic system based on host-guest chemistry [17, 18]. We briefly describe multi-valency effects in the case of polymers functionalized with many ligands.

We consider a flexible polymer with a contour length much larger than the persistence length. Ligands are randomly attached along the polymer chain (see Figure 3.6A). Similar to the nano-particles case above, a reasonable first assumption is that, due to polymer-chain flexibility, all $k$ ligands on a polymer can bind to any of the $n_R$ receptors within a domain on the surface with lateral dimensions comparable to those of the polymer. For simplicity, we describe the surface as a square lattice. The cells of the lattice have linear dimensions comparable to the radius of gyration $R_g$ of the polymer. As in the case of soft multivalent particles, any ligand on the polymer can bind to any receptor in one (and only one) lattice cell, see Figure 3.6. The model is expected to offer a faithful description of the real system if the mean distance between ligands is larger than the Kuhn segment length such that even consecutive ligands along the polymer chain can be treated as uncorrelated.

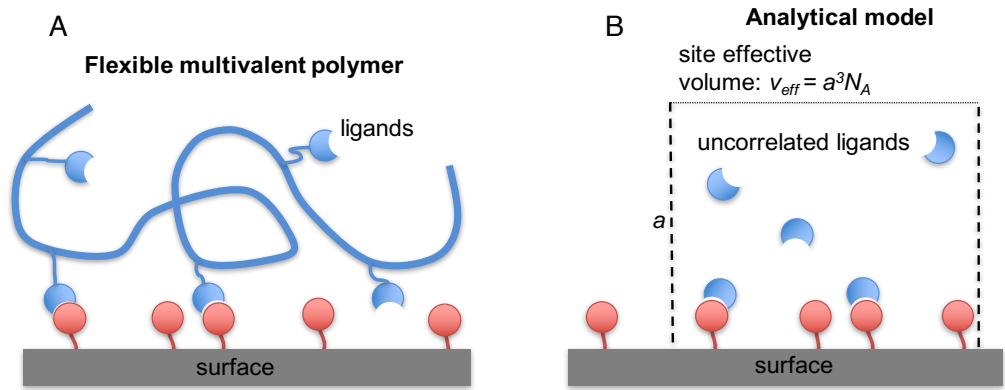

**Figure 3.6** Cartoon of the multivalent polymer model. **A)** Flexible multivalent polymer close to the receptor decorated surface is modelled as **B)** uncorrelated ligands within a lattice site with volume $v_{eff} = \frac{K_A}{K_{intra}} = a^3 N_A$ and $a$ the linear lattice size. The ligands can move and bind to receptors independently within the lattice site, but cannot escape the site individually.

**Multivalent polymer: a cloud of ideal ligands**

The calculation of the avidity constant, via Eq. (3.7), is the same for multivalent polymers or particles. In the case of flexible polymers we can also estimate the intra association constant as $K_{intra} = K_A c_{eff}$, with the effective concentration $c_{eff} \approx 1/N_A a^3$, the lattice size $a = R_g(4\pi/3)^{1/3}$ and $R_g$ the polymer radius of gyration, $N_A$ is the Avogadro's number. This model (and the choice of effective concentration) effectively describes a multivalent polymer as a cloud of ideal gas ligands. Ligands are uncorrelated (can bind independently) but must stay within a lattice site with volume $a^3$, see Figure 3.6. The number of receptors that a polymer can see is then $n_R = \Gamma N_A a^2$, where $\Gamma$ denotes the molar surface density of receptors.

Using the above definitions, we find the following expression for the avidity constant of a multivalent polymer:

$$K_A^{av} = a^3 N_A \left[ \left(1 + \frac{\Gamma K_A}{a} e^{-\beta U_{poly}}\right)^k - 1 \right], \quad (3.11)$$

which is the equation used to obtain adsorption profiles in Figure 3.7. We have added a correction term $U_{poly}$ which takes into account the deviation of the real system to our 'cloud of ideal ligands' approximation. This approximation neglects the polymeric degrees of freedom and, consequently, any spatial correlations between ligands. Moreover, we ignore the fact that the binding free energy of ligands to receptors is changed by the coupling of the ligands to the polymer backbone. These approximations will result in an error of order $k_B T$ and we expect $U_{poly}$ to be $\mathcal{O}(k_B T)$.

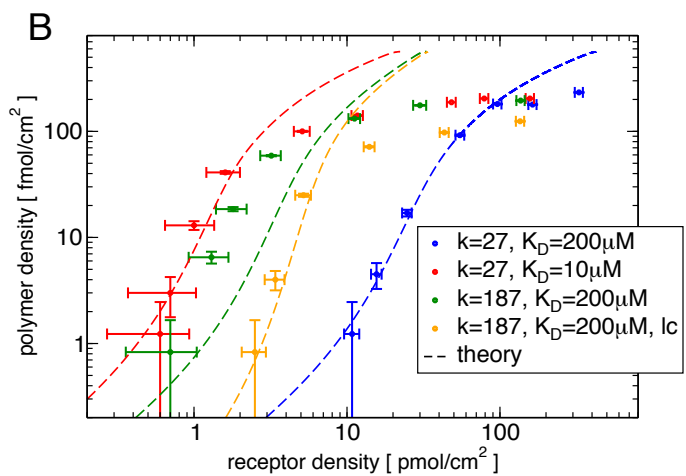

**Figure 3.7** Multivalent polymer adsorption. Experimental adsorption profiles (points with error bars) for hyaluronic acid polymers functionalised with $\beta$-cyclodextrin hosts (HA-$\beta$-CD) binding to surface attached adamantene (affinity $K_D = 1/K_A = 10\mu M$) or ferrocene ($K_D = 200\mu M$) guests reproduced from Refs. [17, 18]. As can be seen, the theoretical adsorption profiles (dashed, dotted or solid lines) match the experimental data well for all valencies ($k$), affinities ($K_D$) and polymer concentration studies. In the Figure legend 'lc' denotes lower concentration of polymers in solution. One parameter ($U_{poly}$ in Eq. (3.11)) was fitted, the value $U_{poly} = 4.6 k_B T$ provides a good fit to all data points.

The analytical expression given by Eq. (3.11) captures the essentials of multivalent polymer adsorption [10]. Importantly the model allows us to predict adsorption profiles and selectivities depending on the physiochemical properties of multivalent polymers, shown in Figure 3.7. Hence, use of the simple theoretical expression given by Eq. (3.11) allows us to design a multivalent polymer such that it will selectively target a desired receptor density. In other words, Eq. (3.11) offers a tool for the rational design of selective targeting.

## 3.4
## Which systems are super selective ?

The discussion thus far focused on selective adsorption of multivalent particles and polymers. We now generalize our treatment and discuss various practical systems. In particular, we will discuss the key role of disorder that is needed to observe super-selective behaviour in multivalent interactions. Specifically, what is needed is that a multivalent entity can bind in many different ways to a receptor-decorated substrate. This kind of disorder is usually not possible for multivalent interactions on the angstrom or nanometer scale, as the interacting units tend to be effectively rigid on that scale. In contrast, larger supramolecular systems (e.g. the binding of a multivalent polymer to a receptor decorated membrane) can sustain the 'disordered' interactions.

### 3.4.1
### Rigid geometry interactions

A prototypical example of multivalent interactions is the fixed (rigid) geometry multivalency shown in Figure 3.8. Two rigid, multivalent entities bind via multiple bonds: as the geometry is rigid, individual bonds either fit together, or they don't. Examples of this kind of interaction include the base pairing between nucleotides in complementary sequences of single-stranded DNA.

Another well-known example of a rigid multivalent interaction is the binding between an enzyme and a substrate. The interaction between a pair of proteins is multivalent, as it involves a number of local interactions of various types (hydrogen bonding, hydrophobic, Van der Waals, electrostatic etc). To a first approximation the enzyme and substrate can be described as rigid objects. This is a simplification as proteins, even in their native state, are not entirely rigid. In any given relative orientation of the ligand to a substrate we find a 2D equivalent of the Figure 3.8. We name this class of multivalent interactions 'rigid geometry multivalency'.

Due to the lack of flexibility of individual bonds, rigid multivalency will generally not show super-selective behaviour. To understand this, consider a simple one-dimensional example of a sequence of rigidly positioned ligands that bind to a commensurate sequence of receptors. One cannot increase the binding site density on the substrate without breaking the commensurability of the binding. Hence, increasing the receptor density will normally decrease the binding strength. In other words: commensurate lock-and-key interactions are not super selective. Interestingly, it seems that the ability of rigid multivalent particles to detect commensurate structures is exploited in nature, for instance in the activation of certain Toll-like receptors [19].

---

10) Ref. [18] also considered the effect of interpenetration of adsorbed polymers. This effect yielded a slightly more complicated theoretical expression. However, the important results, scaling relations and design guidelines, are fully captured by Eq. (3.11).

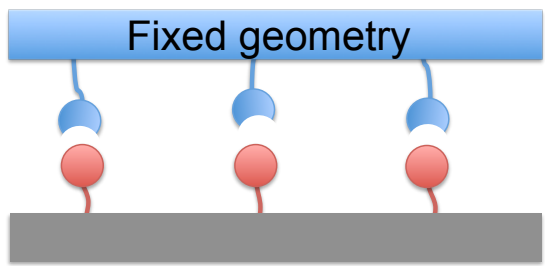

**Figure 3.8** Rigid geometry multivalency. The cartoon presents a prototypical fixed geometry interaction where bonds are commensurate (e.g. DNA base pairing or enzyme-substrate interactions). Such systems generally do not exhibit super-selective behaviour as we cannot increase the (binding site) density on one multivalent entity (substrate) without breaking the commensurability of the bonds. See also Eq. (3.12) and the discussion in the corresponding box.

> **Rigid geometry (commensurate) multivalency can be super selective, but usually is not**
>
> The simplest mean-field model for the commensurable binding case (Figure 3.8) is that every bond pair is equivalent and can be either formed (weight $e^{-f/k_B T}$) or not (weight 1), and all $l$ bond pairs are independent. The avidity constant $K_A^{av\,fix}$ of the multivalent interaction is proportional to the bound partition function $q_b^{fix}$ taking into account all possible states
>
> $$K_A^{av\,fix} \propto q_b^{fix} \approx \left(1 + e^{-f/k_B T}\right)^l \quad . \tag{3.12}$$
>
> Evidently the avidity constant is very sensitive to the number of possible bond pairs $l$, the temperature $T$ and the individual bond strength $f$. The number of possible pairs $l$ depends on the geometry of the interaction. In the simplest model the number of pairs is given by $l = \min[n_R, k]$, limited by whichever substrate or the multivalent entity has a smaller number of sites. [8]. Hence, rigid geometry multivalent interactions can show super-selective behaviour, but only when the multivalent construct initially had an excess number of binding sites compared to the substrate. Furthermore, when increasing the number of binding sites on the substrate, geometric constraints (commensurability) must be obeyed.

**14**

### 3.4.2
### Disordered multivalency

Super-selective behaviour can be exhibited by multivalent systems that can increase the number of possible bonds as the density of receptors increases. As we saw above, fully ordered multivalent systems only bind optimally to commensurate receptor arrangements. To achieve super selectivity, we typically need some kind of disorder or randomness in the geometry of binding. The ability to increase the number of bonds with increasing receptor density can be due to: (i) long, flexible binders, (ii) mobile receptors, or (iii) random binder positions. Figure 3.9 shows schematic examples of these three cases. Different types of bond disorder will result in different expressions for the bound partition functions (and therefore, for the avidity constants), see Eqs. (3.13 - 3.15). However, they all show similar super-selective behaviour (see Figure 3.10).

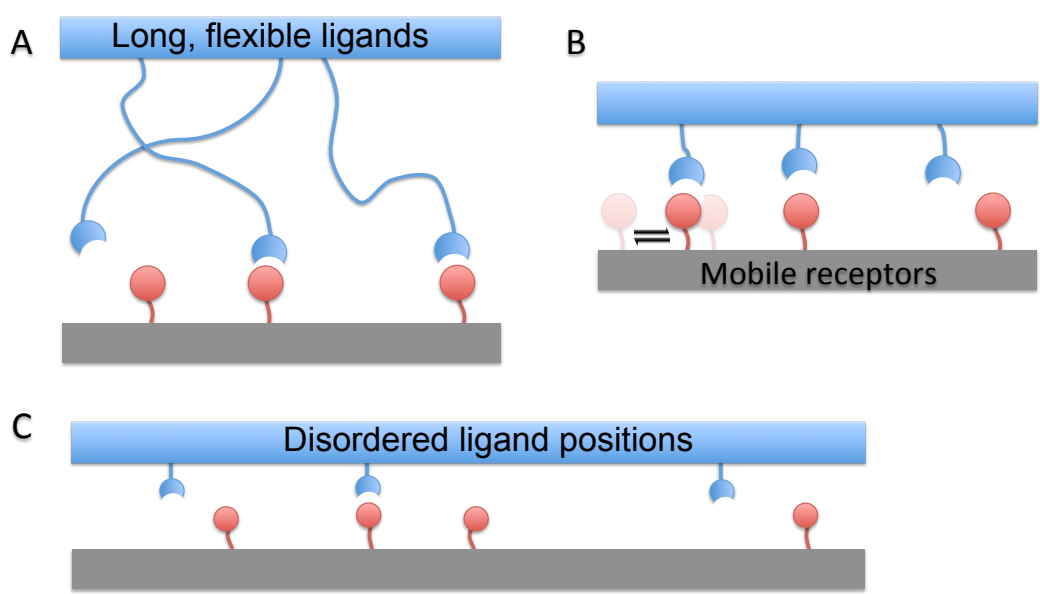

**Figure 3.9** Disordered multivalent systems. Three characteristic types of multivalent interactions are shown: **A**) long, flexible binders, **B**) mobile receptors, **C**) disordered, random positions of individual binders. Different types can behave slightly differently, see the adsorption profiles in Figure 3.10. However, they all exhibit super selectivity, and consequently, any practical system that is similar to at least one of them, will be super selective.

---

**Disordered systems: different, yet similar**

Different forms of disorder may cause super-selective behaviour in multivalent systems. The theoretical expressions for the partition function (and hence the avidity constant) of the bound state will depend on the nature of the disorder. Below, we list a few examples:

> - Long flexible ligands, Figure 3.9A); the number of ligands and receptors is fixed and all $k$ ligands can reach any of the $n_R$ receptors
>
> $$q_b(n_R, k) = \sum_{i=1}^{\min(n_R,k)} \binom{n_R}{i}\binom{k}{i} i!\, e^{-\beta i f}, \qquad (3.13)$$
>
> which is the expression that we have already used above (see Eqs. (3.5,3.6)).
> - Mobile receptors (Figure 3.9B)); the number $n_R$ of accessible receptors fluctuates and is Poisson distributed with mean $\tilde{n}_R$. Poisson averaging of Eq. (3.13) over $n_R$, we find
>
> $$q_b(\tilde{n}_R, k) = \left(1 + \tilde{n}_R e^{-\beta f}\right)^k - 1, \qquad (3.14)$$
>
> which is the same as expressions (3.7, 3.9) already considered above.
> - Large colloids (or cells) with disordered or mobile ligand positions (Figure 3.9C)); both the number of ligands and the number of receptors are Poisson distributed with mean $\tilde{k}$ and $\tilde{n}_R$ respectively. Poisson averaging Eq. (3.14) over $k$, we find
>
> $$q_b(\tilde{n}_R, \tilde{k}) = e^{\tilde{n}_R \tilde{k} e^{-\beta f}} - 1. \qquad (3.15)$$
>
> A comparison between the predicted behaviour of these different systems is shown in Figure 3.10. In the limit of high valency ($k \gg 1$, $n_R \gg 1$) and weak bonds ($n_R e^{-\beta f} < 1$, $k e^{-\beta f} < 1$) the behaviour of all systems converges to the same form.

At first sight, it would seem that the case of mobile receptors shown in Figure 3.9B) should be rather different from the immobile case. However, since the receptors are mobile, each ligand can, in principle, bind to any receptor. In this light the two problems become very similar. Another way of looking at the system with mobile receptors is to consider the receptors as a (two-dimensional) 'ideal gas' of particles that can bind to the ligands with an interaction strength $f$. Up to a concentration-independent term $\mu_R^0$, the chemical potential of these receptors is given by $\mu_R \approx k_B T \log(n_R)$. A small change in the receptor concentration $n_R$ leads to a small change in the chemical potential $\mu_R$, which alters the probability of each and every individual ligand binding. For multivalent particles a small change per ligand adds up to a large change per particle [11]. Clearly, the binding probability depends on $n_R$, see Refs. [19, 23] for practical examples of super selectivity with mobile receptors. We note that for dilute receptors the chemical potential is dominated by the translational entropy. Hence, the origin of super selectivity is entropic, also for mobile receptors.

---

11) If we assume that there are many more receptors than ligands, we can then write the bound-state partition function for $k$ ligands as $q_b \approx (1 + c n_R e^{-\beta f})^k - 1$, where the constant $c$ depends only on the concentration-independent part of the chemical potential $\mu_R$ as $c = -k_B T \ln \mu_R^0$. In the case of many weak binders: $c n_R e^{-\beta f} \ll 1$ and we can approximate $q_b \approx e^{k c n_R e^{-\beta f}} - 1$.

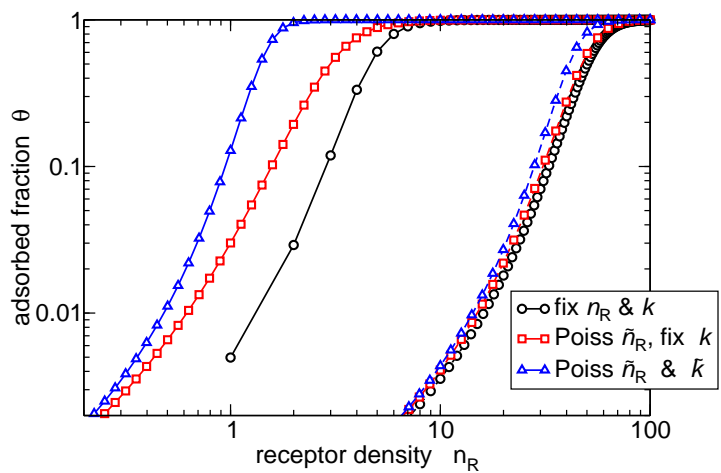

**Figure 3.10** Adsorption profile for different disordered systems, depicted in Figure 3.9. Different systems show qualitatively similar super-selective behaviour. For large number of bonds the adsorption profiles converge. The plots shown were generated using the Langmuir expression for the adsorption isotherm Eq. (3.8) with activity $z = 0.001$. We used expressions (3.13, 3.14, 3.15) with $k = 5$ and $\beta f = 0$ to compute adsorption isotherm in the case of a few strong bonds (solid lines). To represent the case of many weak bonds (dashed lines), we used the same equations but assumed $k = 25$ $\beta f = 5$.

Finally, for immobile randomly distributed binders shown in Figure 3.9C) the intuitive reasoning for super selectivity follows from our initial discussion in the introduction. Let us consider two ligand/receptor-decorated multivalent nanoparticles, $A$ and $B$ that can attach through ligand-receptor binding. The binding moieties are randomly distributed on both nanoparticles. From a point of view of a particular ligand on particle $A$, the probability of it binding, $p_{1A}$, is to a first approximation linear in the density $n_R$ of complementary receptors on particle $B$. The number of possible bonds in the contact area is proportional to the number of ligands $k$ in that area. The net result is that the binding probability depends exponentially on the product of $k$ and $n_R$, as would follow from Eq. (3.15).

We note that in the cases of fixed short ligands we have only illustrated and discussed the two limiting cases: (i) perfectly complementary rigid interaction (Figure 3.8) and (ii) disordered interaction case (Figure 3.9C). Practical systems will fall between these two extremes. As a rule of thumb, small molecules and macromolecules, such as DNA or proteins, or virus capsids have a rather well defined geometry and we expect their interactions to be closer to the rigid geometry case. On the other hand, the spatial distribution of binders (ligands) on entities larger than a few nano-metres is, in general, more disordered; be they man-made such as DNA coated colloids [24, 25, 26], or natural such as cells.

We have presented simple analytical models that can be used to rationalise and understand super selectivity in various multivalent systems. In the case of polymers, the simple model works very well (see 3.7). However, certain systems have been studied in a greater detail. For these cases, more sophisticated (and more complex) models have been developed. For example, cell endocytosis of a virus is mediated by a multivalent interaction between membrane proteins (receptors) and virus capsid proteins (ligands). But to model the process, one should account for membrane elasticity and, in some cases, also for active processes [27]. More detailed models of multivalent polymer adsorption have recently been developed [28, 29]. A theory of valence-limited interactions explicitly taking into account specific positions and different types of tethered binders requires the self-consistent solution of a system of equations [30, 31], the framework was also extended to mobile ligands [32]. A complementary approach is based on a saddle-point approximation for the binding free energy [33]. We note that the results presented in these papers support the conclusions about super-selective behaviour that we have obtained here using much simpler models.

## 3.5
## Design principles for super-selective targeting

Clearly, super-selective targeting has important practical applications (as even viruses seem to 'know'). It is therefore important to formulate design principles for achieving optimal super selectivity. To formulate design rules, we start once again from the simple model described above: multivalent particle docking to a receptor-decorated surface (e.g. a cell). The density of receptors on the surface is again measured by $n_R$, the mean number of receptors in the contact area (i.e. the area accessible to a docked particle). In many cases of practical interest, we aim to target only those surfaces (e.g. a cell surface) that have a receptor concentration above a certain threshold. How should we design the particle to target this surface optimally? Our control parameters are the valency $k$, the ligand-receptor binding strength $f$, and the activity of particles in solution $z$.

> **Optimising the selectivity**
>
> In terms of the theoretical expressions, Eqs. (3.8) and (3.9), we aim to maximise the selectivity
>
> $$\alpha(n_R) = \frac{\partial \log \theta}{\partial \log n_R} \tag{3.16}$$
>
> at a given desired receptor density $n_R$. We note that partition function $q_b$ (Eq. (3.9)) and its derivative are increasing functions of $n_R$, $k$ and $-f$. Hence, we expect the selectivity (slope) to be the highest just before denominator in Eq. (3.8) becomes important and the maximal selectivity will be found when $zq_b \approx 1$. Using Eq. (3.9) we can solve this equation, which yields a relation between $k$ and $f$
>
> $$k = \frac{-\log(z)}{\log\left(1 + n_R e^{-\beta f}\right)}. \tag{3.17}$$

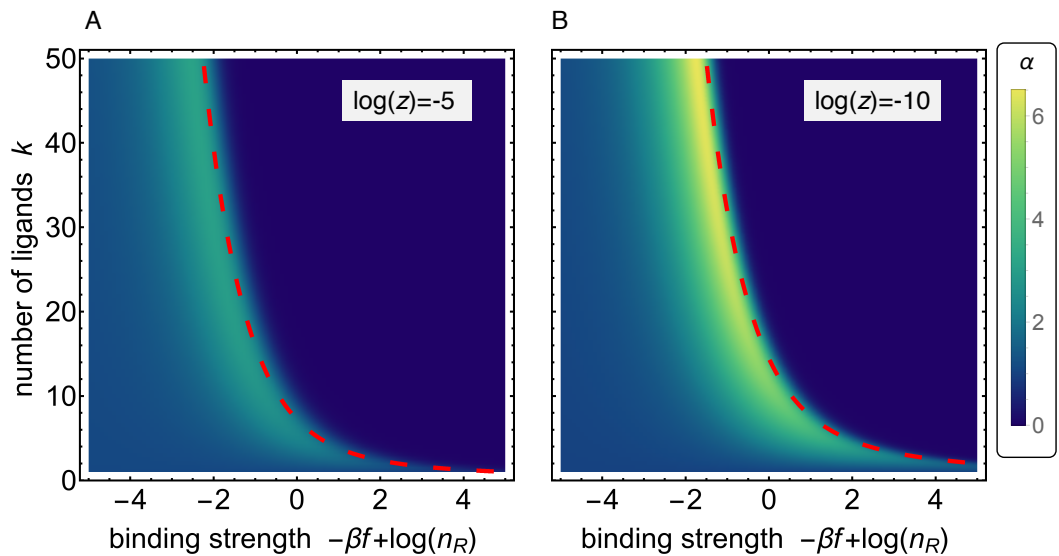

**Figure 3.11** The selectivity landscape as function of the valency $k$ and the rescaled binding strength $-\beta f + \log(n_R)$. The landscape was obtained by calculating the selectivity $\alpha$ using Eq. (3.16). The activity of multivalent particles was chosen as: A) $z = \exp(-5)$, and B) $z = \exp(-10)$. Both plots use the same colour scale. The dashed curves represent the approximate optimal selectivity relation given by Eq. (3.17), which rather accurately fits the maximum selectivity region.

When $zq \approx 1$ we also have approximately $\theta \approx \frac{1}{2}zq$ and the selectivity becomes

$$\alpha \approx k\frac{n_R e^{-\beta f}}{1 + n_R e^{-\beta f}} = -\log(z)\frac{n_R e^{-\beta f}}{(1 + n_R e^{-\beta f})\log(1 + n_R e^{-\beta f})}, \quad (3.18)$$

where, in the last step, we used Eq. (3.17).

Expanding the above function to first order for weak/strong binding we find the characteristic behaviour: (i) In the case of strong binding the selectivity is

$$\alpha(n_R e^{-\beta f} > 1) \approx \frac{-\log(z)}{-\beta f + \log(n_R)}, \quad (3.19)$$

and in the weak binding limit

$$\alpha(n_R e^{-\beta f} < 1) \approx -\log(z). \quad (3.20)$$

Clearly, the selectivity is maximal in the weak-binding limit and is determined by the logarithm of the activity, see landscape plots in Figure 3.11. In the strong-binding limit, the selectivity decreases with increasing strength of the individual bonds. We remember that $z = \rho \frac{K_A}{K_{intra}}$ and $e^{-\beta f} = K_{intra}$.

The landscape plots of selectivity as a function of the valency $k$ and bond strength $f$ are shown in Figure 3.11. We immediately notice three features: (i) High selectivity appears only in a small region of the parameter space, along the curve predicted by Eq. (3.17). (ii) The selectivity reaches a plateau value at large valencies $k$ and weak individual bonds. (iii) Maximum selectivity is limited by the activity $z$; lowering the activity (or density) of multivalent particles yields a higher selectivity.

The dimensionless activity $z = \rho \frac{K_A}{K_{intra}}$ depends on the density $\rho$, but also on the ratio of the equilibrium constants for the formation of the first bond, and for the formation of subsequent ligand-receptor bonds in a particle-substrate complex (see Eq. 3.5). Therefore, even at large densities, selectivity can be substantial if the ratio $\frac{K_A}{K_{intra}}$ is small. This can be achieved by adding a non-specific repulsion between the multivalent entities (for instance, by coating the particle with inert polymer that provides steric repulsion [34]). Such a repulsion would present a barrier to particle association but would not prevent additional bonds from forming once the barrier is overcome: the result would be a reduction in $K_A$ due to repulsion, but as $K_{intra}$ would be less affected, this steric repulsion would decrease the ratio $\frac{K_A}{K_{intra}}$.

Our calculations show that selectivity is suboptimal when using high affinity bonds. However, strong affinity multivalent constructs can still behave super selectively ($\alpha > 1$) if their activity (concentration) in the solution is low enough, see Eq. (3.19). This suggests that, although in principle it is possible to design a super-selective system based on very strong affinity interactions, such as the biotin-streptavidin pair, such a system would only be super selective at extremely low concentrations where the kinetics would be too slow for practical applications.

Multivalency leads to super selectivity, but it also leads to high sensitivity of binding to the variation in other relevant quantities. Therefore, in practical applications, it is important to control (or, at least know) parameters such as temperature, pH, ionic binding strength when using multivalent particles for selective targeting. The parameter range that yields high selectivity is rather small, see Figure 3.11B). A brute-force 'random' search in design-parameter space is, therefore, unlikely to find the optimal selectivity region. We hope that the theoretical guidelines and design principles set forth in this chapter will enable a more rational design of particles for super-selective targeting.

We condense the results shown in Figure 3.11 and our theoretical considerations, Eq. (3.18), in a set of simple design rules for multivalent binding that yield maximum selectivity. We use our dimensionless statistical mechanics notation, which can be straightforwardly converted to chemical equilibrium units using $z = \rho \frac{K_A}{K_{intra}}$ and $e^{-\beta f} = K_{intra}$, as discussed in the Notation box.

1) The maximal possible selectivity $\alpha$ is limited by the activity of multivalent particles in solution: $\alpha_{max} \; - \log(z)$ so the activity $z$ of multivalent binders should be small.
2) Many weak bonds are better than few strong ones. The selectivity is also limited by the valency $k$, until a point of saturation given by $k \sim -\log(z)$. The first two design rules together state that the maximal selectivity is limited by either the valency $k$ or the $-\log(z)$, whichever is smaller.
3) the relationship between the ligand number $k$ and binding strength $f$ should be obeyed: $k = \frac{-\log(z)}{\log(1+n_R e^{-\beta f})}$. Together with the above rule, this one states that to achieve maximal selectivity individual bonds should be very weak $K_{intra} = e^{-\beta f} < 1/n_R$. In other words, the fraction of bound receptors/ligands should always remain small.

The main assumptions used to arrive at these design rules are: (i) ligands are identical and bind independently, (ii) all ligands of a (surface bound) multivalent construct can reach all surface attached receptors within a lattice site, but cannot bind to any receptor outside of the site (see Figure 3.6). (iii) Receptors, ligands or particles have no interactions except for the steric repulsion and ligand-receptor affinity.

## 3.6
## Summary: it is interesting, but is it useful ?

We have shown that weak, multivalent interactions can result in a super-selective behaviour where the overall interaction strength becomes very sensitive to the concentration of individual binders (receptors). We presented a simple yet powerful analytical model with good predictive power for designing multivalent interactions. We expect that, even in cases where the simple model fails quantitatively, the above design rules will still provide a good starting point for designing super selectivity in practical multivalent systems. Figure 3.12 summarises advantages of weak multivalent interactions in selective targeting.

We can imagine effective purification devices where nano objects of different valencies are passed through super-selective sieves. In the field of material self-assembly, multivalent supramolecular entities could be designed to hierarchically assemble depending on the valency, thus enhancing the precision of self-assembled constructs [25].

The ability to target diseased cells pathogens based on the surface concentration of certain (over)expressed receptors would be of huge practical importance. At present, the delivery of pharmaceutical compounds to specific cells is usually based on the existence of a specific marker (e.g. a sugar or a peptide fragment) that is unique to the targeted cell type. The current wisdom seems to be to functionalise drugs or drug carriers such that they bind strongly to the specific marker. This strategy is fine if the target cells (e.g. bacteria) are very different from the cells of the host, and carry very different markers.

However, the strong-binding strategy becomes problematic if one wishes to target, say, cancerous cells, which are usually very similar to our healthy cells. Cancerous cells typically over-express markers that are also present, be it in smaller quantities, on healthy cell surfaces. Examples are the CD44 ('don't eat me' receptor) or the folic receptor. In such cases, a compound that binds strongly to the over-expressed marker will also bind to (and kill) healthy cells. The insensitivity of strong binders to the surface concentration of markers is one of the main reasons why antibiotics can be efficient with few side effects (in most patients), while chemotherapy is directly harmful to our body.

As outlined in this chapter, carefully designed multivalent drugs could be targeted

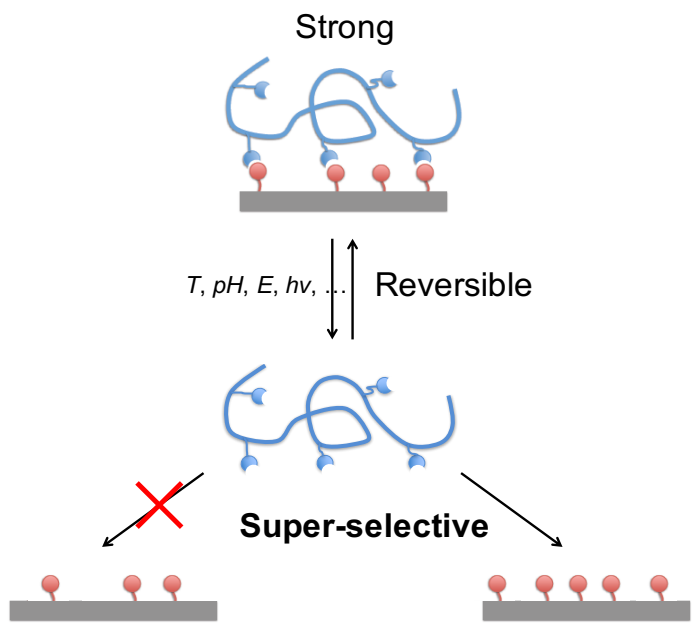

**Figure 3.12** Advantages of using weak bonds. Contrary to strong monovalent antibody-antigen interactions and covalent bonds, multiple weak complexes can be disassembled (one by one) using different environmental stimuli (temperature, interaction strength, pH, light), which provides flexibility and reversibility. Examples of systems that exploit multivalency are dendrimers [35], stimulus-responsive coatings [36], renewable sensors for biomolecules [37], reversible gels [38] and gel-particle glue [39]. Importantly, external stimuli can be used to tune the super-selectivity region to the desired surface density of receptors. For example, one could exploit the acidic extracellular environment of tumour tissues to improve the efficiency of drug targeting using multivalent particles.

super selectively only to cells with cognate receptor concentration above a certain threshold value [8, 40, 41]. Furthermore, in a living cell, receptor interactions and signalling play a major role which can further enhance the non-linear response of the system [42, 43, 44, 45, 46]

Multivalency extends the sensitivity of interactions into the receptor density domain. Moreover, it enables the design of specific, highly selective interactions based on the concentration of ligands or binders, as well as on their chemical nature, thus opening up the possibility for selective targeting with minimal side effects.

## 3.7
## Appendix: What is effective molarity ?

Effective molarity ($EM$) is an empirical concept that is commonly used to relate the kinetics and equilibria of intramolecular and intermolecular reactions [9, 11, 10]. It is defined as

$$EM = \frac{K_A{}^{intra}}{K_A{}^{inter}} , \qquad (3.21)$$

where $K_A{}^{intra}$ and $K_A{}^{inter}$ are the equilibrium association constants. $EM$ has units of molar concentration and is a useful measure of multivalent interactions efficacy, see Figure 3.13. For example, when the concentration $\rho$ of multivalent ligands in solution is high $\rho \gg EM$ multivalent effects are suppressed and ligands will bind monovalently. On the other hand when $\rho \ll EM$ multivalent interactions dominate over monovalent binding. Additionally, $EM$ allows us to de-convolute the intra equilibrium constant into a simple part ($K_A$) due to bond formation, and a complicated part ($EM$) related to the change of conformational entropy and free energy upon binding, see Refs. [9, 10, 12, 13] for more discussion.

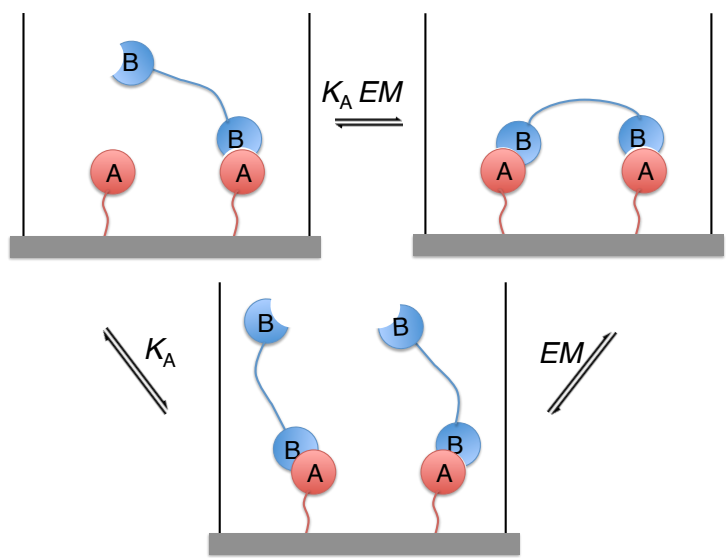

**Figure 3.13** The concept of effective molarity. The above cycle shows the 3 different states that divalent ligands ($BB$) can bind to two receptors ($AA$) (unbound state is omitted). We have 3 distinct states and, therefore, need 2 equilibrium constants to characterise the equilibrium properties of the system, $K_A$ and $EM$. A product of the two is often called an intra association constant $K_A^{intra} = K_A EM$. A useful reference point is that for a divalent ligand/receptor system and saturated receptors, $EM$ determines the concentration of divalent ligands $[BB]$ in solution at which we expect equal number of singly and doubly bonded ligands.

However, it is important not to over-interpret the meaning of 'effective' concentrations. The name suggests that we can calculate the internal chemical equilibria of multivalent interactions simply by using some effective concentrations of ligands. That, however, is not quite the case, as the expressions for association equilibrium between two compounds do not carry over to the situation when the numbers involved are small.

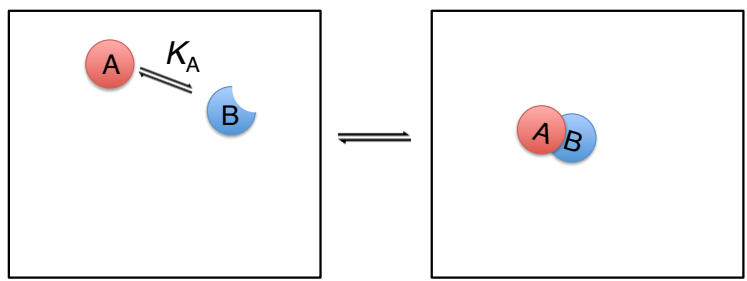

**Figure 3.14** Dimerisation reaction in a small box. We have two particles (a single $A$-type and a single $B$-type) in a box with volume $V$. We assume that, although the particles can bind, they do otherwise behave as an ideal gas. We wish to calculate the relation between probability of dimerisation and equilibrium association constant $K_A$. Simply calculating effective concentrations of $[A]$, $[B]$ and $[AB]$, and using standard chemical equilibrium equation $\frac{[AB]}{[A][B]} = K_A$ gives a wrong answer, see boxes on dimer reactions.

Let us consider a prototypical system: Only two particles (ligands) in a box with volume $V$. The particles can associate with an equilibrium constant $K_A$ that was predetermined for us, see Figure 3.14. We wish to calculate the association probability of these two particles. To obtain the correct result we can calculate the partition functions of the bound/unbound state.

---

**Dimerisation: Correct calculation**

The unbound partition function of two molecules in the box is

$$q_u = V^2 , \tag{3.22}$$

since we assume both particles are non-interacting and can independently explore the entire box volume $V$. The bound partition function is

$$q_b = V v_0 e^{-\beta \Delta G} , \tag{3.23}$$

with $e^{-\beta \Delta G} = K_A \rho_0$ the dimerisation free energy and $v_0 = \frac{1}{\rho_0 N_A}$ the microscopic volume of the bond and $\rho_0 = 1\text{M}$ the standard concentration. The ratio of the partition functions determines the probability that a dimer is formed.

$$\frac{1 - p_u}{p_u} = \frac{q_b}{q_u} = \frac{K_A}{V N_A} , \tag{3.24}$$

with $p_u$ denoting the unbound probability and the probability that two particles are bound is simply $p_b = 1 - p_u$.

On the other hand, if we naively make use of the expression for chemical equilibrium in a bulk mixture binary chemical equilibrium, we do not reproduce the correct result.

> **Dimerisation: Wrong calculation**
>
> We could simply rationalise that the effective (time averaged) concentration of unbound chemicals is
>
> $$[A] = [B] = \frac{p_u}{V N_A} \; , \tag{3.25}$$
>
> where $p_u$ is the probability that $A$ and $B$ are unbound, $V$ is the box volume and we have added the Avogadro's number $N_A$ to make $[A]$ and $[B]$ a molar concentration. Similarly for the dimerised state $[AB] = \frac{1-p_u}{V N_A}$. Hence, in line with standard chemical dimerisation reaction, we could reason that
>
> $$K_A = \frac{[AB]}{[A][B]} = \frac{1-p_u}{p_u^2} V N_A \; , \tag{3.26}$$
>
> which is clearly different from Eq. (3.24)

Treating the system as a bulk binary reaction is not valid for only two dimerising particles. The approach is valid in the thermodynamic limit where the chemical potential of a molecular species can be related to the logarithm of its concentration. What it boils down to is that Stirling's approximation is valid only for large number of particles $\log N! \approx N \log N - N$, it is clearly wrong when $N$ equals 1 or 2. The same problem occurs when trying to calculate equilibrium constant from molecular dynamics simulations using small system sizes [47]

The above example might seem rather abstract. However, it exposes a potential pitfall of misusing 'effective' concentrations. The same pitfall is encountered when calculating binding probabilities of multivalent ligands, because the reactions shown in Figures 3.14 and 3.13 are very similar. For example, one could naively argue that both the unbound ligand ($A$) and receptor ($B$) in Figure 3.13 are flexible and can explore some effective volume $V$ and have some effective concentration within this volume. One then applies a 'Local chemical equilibrium' (LCE) assumption [24, 48] which, in our simple system is given by Eqs. (3.25, 3.26). But this procedure does not generally give a correct result. It becomes a good approximation only in the limit of weak binding[12] or a very large valency where the Stirling's approximation becomes applicable.

It should be clear that effective molarity is not really a concentration[13]. Rather, it is a quantity with the dimensions of concentration, defined by Eq. (3.21). We can view the effective molarity as a measure for the probability that an unbound ligand and receptor would overlap in space (and hence come into position to bind). In an idealised system, neglecting the effects of the linker and orientational correlations in the unbound state, this probability is related to an effective concentration of, say, a ligand ($B$) as experienced by its complementary receptor ($A$) [9, 12, 14]. This is exactly the 'cloud of ideal ligands' approximation we have used as a starting point for our theory of multivalent polymer adsorption, Eq. (3.11).

---

12) for weak binding $p_u \approx 1$ and Eq. (3.26) becomes a very good approximation to Eq. (3.24)
13) The effective molarity can be calculated via relative concentrations of singly and doubly bound states in solution, see Figure 3.13, but *EM* as such is determined only by the interaction between the two multivalent binders.

In the case of our simplified system of 2 dimerizing particles (Figure 3.14) the effective concentration $c_{eff}$ of type-$A$, as experienced by type-$B$, (or vice versa) is

$$c_{eff} = 1/(VN_A) , \tag{3.27}$$

where we recall that $V$ is the box volume. We can think of particle $A$ adsorbing to particle $B$ and the ratio of probabilities of being bound to unbound becomes

$$p_b = K_A c_{eff} p_u , \tag{3.28}$$

which is consistent with the correct result, Eq. (3.24). We could view $c_{eff}p_u$ as the concentration of unbound $A$.

Applying this concept to dimer adsorption (Figure 3.13) we would find that the empirically calculated effective molarity (Eq. (3.21)) is similar to the theoretical effective concentration $EM \sim c_{eff}$ (in our idealised system they are equal). Therefore, effective concentration, when applied properly, is a useful concept when attempting to theoretically predict equilibria of multivalent binding.

**Acknowledgements**

We thank Galina V. Dubacheva and Stefano Angioletti-Uberti for useful suggestions to the manuscript and help with designing figures. This work on multivalency was supported by the ERC Advanced Grant 227758 (COLSTRUCTION), ITN grant 234810 (COMPPLOIDS) and by EPSRC Programme Grant EP/I001352/1. T.C. acknowledges support form the Herchel Smith fund.